\documentclass{emulateapj}

\usepackage{apjfonts}

\newcommand{\src}{RRAT~J1819--1458}

\newcommand{\srcx}{CXOU~J181934.1--145804}

\catcode`\@=11
\newcommand{\gapprox}{\mathrel{\mathpalette\@versim>}}
\newcommand{\lapprox}{\mathrel{\mathpalette\@versim<}}
\newcommand{\propapprox}{\mathrel{\mathpalette\@versim\propto}}
\newcommand{\@versim}[2]
  {\lower3.1truept\vbox{\baselineskip0pt\lineskip0.5truept
\ialign{$\m@th#1\hfil##\hfil$\crcr#2\crcr\sim\crcr}}}
\catcode`\@=12

\shorttitle{X-RAY COUNTERPART TO RADIO TRANSIENT J1819--1458}
\shortauthors{REYNOLDS ET AL.}

\begin{document}

\title{Discovery of the X-ray Counterpart to the \\
Rotating Radio Transient J1819--1458}


\author{Stephen P. Reynolds,\altaffilmark{1} Kazimierz J.
Borkowski,\altaffilmark{1} Bryan M. Gaensler,\altaffilmark{2,3}
Nanda Rea,\altaffilmark{4} Maura McLaughlin,\altaffilmark{5} \\
Andrea Possenti,\altaffilmark{6} Gianluca Israel,\altaffilmark{7}
Marta Burgay,\altaffilmark{6} Fernando Camilo,\altaffilmark{8} 
Shami Chatterjee,\altaffilmark{2,9} Michael Kramer,\altaffilmark{5} 
Andrew Lyne,\altaffilmark{5} and Ingrid Stairs\altaffilmark{10}}

\altaffiltext{1}{Physics Department, North Carolina State University,
    Raleigh, NC 27695}
\altaffiltext{2}{Harvard-Smithsonian Center for Astrophysics, 60 Garden
Street, Cambridge, MA 02138}
\altaffiltext{3}{Alfred P.\ Sloan Research Fellow}
\altaffiltext{4}{SRON -- Netherlands Institute for Space Research, 
   Sorbonnelaan, 2, 3584 CA, Utrecht, The Netherlands}
\altaffiltext{5}{Jodrell Bank Observatory, University of Manchester,
   Macclesfield, Cheshire SK11 9DL, UK}
\altaffiltext{6}{INAF -- Osservatorio Astronomico di Cagliari,
Loc. Poggio dei Pini, Strada 54, 09012 Capoterra, Italy}
\altaffiltext{7}{INAF -- Osservatorio Astronomico di Roma,
via Frascati 33, I-00040 Monteporzio Catone, Italy}
\altaffiltext{8}{Columbia Astrophysics Laboratory, Columbia University, 550
West 120th Street,
New York, NY 10027}
\altaffiltext{9}{Jansky Fellow, National Radio Astronomy Observatory}
\altaffiltext{10}{Department of Physics and Astronomy, University of British 
Columbia, Vancouver, BC V6T 1Z1 Canada}

\vskip 1 truein

\newpage

\begin{abstract}

We present the discovery of the first X-ray counterpart to a Rotating
RAdio Transient (RRAT) source.  \src\ is a relatively highly
magnetized (B $\sim 5\times10^{13}$\,G) member of a new class of
unusual pulsar-like objects discovered by their bursting activity at
radio wavelengths. The position of \src\ was serendipitously observed
by the {\sl Chandra} ACIS-I camera in 2005 May.  At that position we
have discovered a pointlike source, \srcx, with a soft spectrum well
fit by an absorbed blackbody with $N_H = 7^{+7}_{-4} \times 10^{21}$
cm$^{-2}$ and temperature $kT=0.12 \pm 0.04$\,keV, having an
unabsorbed flux of $\sim2 \times 10^{-12}$\,ergs\,cm$^{-2}$\,s$^{-1}$
between 0.5 and 8 keV.  No optical or infrared (IR) counterparts are
visible within $1''$ of our X-ray position.  The positional
coincidence, spectral properties, and lack of an optical/IR
counterpart make it highly likely that \srcx\ is a neutron star and is
the same object as \src.  The source showed no variability on any
timescale from the pulse period of 4.26~s up to the five-day window
covered by the observations, although our limits (especially for
pulsations) are not particularly constraining.  The X-ray properties
of \srcx, while not yet measured to high precision, are similar to
those of comparably-aged radio pulsars and are consistent with thermal
emission from a cooling neutron star.

\end{abstract}


\section{Introduction}
\label{intro}

The discovery of a new class of ``Rotating RAdio Transients'' (RRATs)
has recently been reported by McLaughlin et al.~(2005).  These
objects, 11 so far identified, are characterized by repeated radio
bursts with durations between 2 and 30 ms and average intervals
between bursts ranging from 4 minutes to 3 hours.  Their dispersion
measures (DMs) place them within the Galactic plane at distances from
2 to 7 kpc.  If bursts are periodic, periods can be found from the
greatest common divisor of the differences between burst arrival
times.  For ten of the sources, this calculation results in periods
between 0.4 and 7 seconds, suggesting that the objects are rotating
neutron stars. The periods measured for the RRATs are longer than
those of most normal radio pulsars and more similar to those of the
populations of X-ray dim isolated neutron stars (XDINSs; Haberl~2004)
and magnetars (Woods \& Thompson~2006).  For the three sources with
the highest bursting rates, period derivatives, $\dot P$, have been
measured.  No binary motion is detected. If the $\dot P$ values are
interpreted as due to magnetic dipole spin-down, they imply
characteristic ages and magnetic field strengths in the general range
of pulsars.

In this paper we report the X-ray detection of \src, the first
detection at other wavelengths of any of the RRATs. This source has a
4.26-s period, a relatively high inferred characteristic surface
dipole magnetic field strength of $5\times10^{13}$~G, a characteristic
age $P/2\dot{P} = 117$~kyr, and a spin-down luminosity of
$3\times10^{32}$~ergs~s$^{-1}$. The distance of this source inferred
from its DM using the electron-density model of Cordes \& Lazio~(2002)
is 3.6~kpc, with considerable uncertainty.  \src\ is characterized by
radio bursts of average duration 3~ms, with one burst detected every
$\sim$~3~minutes. This object was fortuitously in the ACIS-I field of
a 30 ks {\sl Chandra} observation toward the Galactic supernova
remnant G15.9+0.2 (Reynolds et al., in preparation).  The brightest
source on any of the 6 CCD chips of the {\sl Chandra} field, besides
G15.9+0.2 itself, is coincident to within $2''$ with the radio
position of \src\ (whose error ellipse has semimajor axes $5'' \times
32'' $).  The positional coincidence, as well as properties we
describe below, make us confident that this new source, which we
designate \srcx, is the X-ray counterpart to the radio-bursting source
\src.

\section{Observations}
\label{obs}

The {\sl Chandra} observations were performed with the ACIS
instrument in full-frame mode on 2005 May 23 (10 ks), May 25 (5 
ks), and May 28 (15 ks).  \srcx\ lies on the I3 chip of the ACIS-I
camera.  We checked aspect correction, created a new level--1 event
file appropriate for VFAINT mode (without applying pixel
randomization in energy), and applied light-curve filtering to
reject flares.  CTI correction was applied and calibration was
performed using CALDB version 3.1.0. 

\begin{figure}
\epsscale{.80}
\plotone{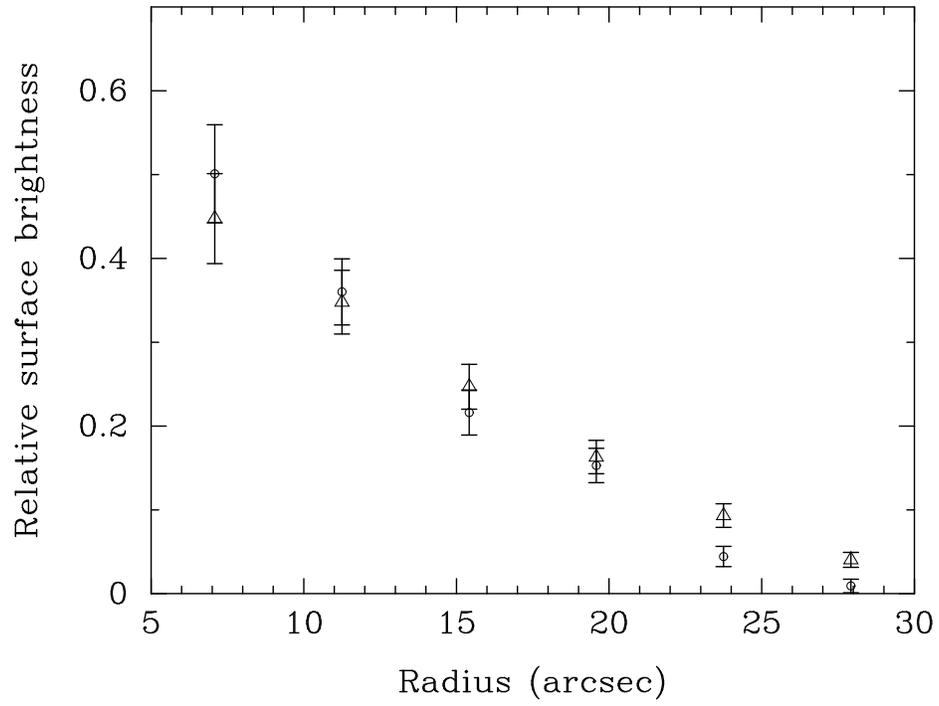}
\caption{The radial profile of \srcx\ at 1~keV (circles) and
the corresponding predicted PSF at 
that location (triangles).  There is no evidence for source
extension.  Errors are statistical only (90\% limits).
\label{profile}}
\end{figure}

\srcx\ is by far the brightest compact source on the I3 chip or on any
of the others.  We find a net count rate after background subtraction
of 0.018 ct s$^{-1}$ (0.5--8 keV), for a total of $524 \pm 24$ counts.
Radial profiles were created for the source and for the 1 keV
point-spread function (PSF) at that location on the I3 chip, and are
shown in Figure~\ref{profile}.  There is no evidence for any extended
emission.  The source position (fit with WAVDETECT) is $18^{\rm
h}19^{\rm m} 34\fs17 \pm 0\fs02,\ -14^\circ 58' 04\farcs6 \pm
0\farcs2$ (J2000).  The errors are statistical only.  We measured the
position of a bright star visible in X-rays to be within $0\farcs1$ of
its USNO UCAC2 position, but position errors for sources $10'$
off-axis may be $ \la 0\farcs4$ (Getman et al.~2005).  We adopt
$0\farcs5$ as a conservative total error estimate.

We searched various catalogs for optical or IR counterparts within
$1''$ of our best-fit position.  No counterparts were detected in the
2MASS catalog or any of the others searched by
VizieR\footnote{http://vizier.u-strasbg.fr/viz-bin/VizieR}, with
magnitude upper limits of $19\fm9$ (B2), $18\fm0$ (R1), $17\fm5$ (I),
$15\fm6$ (J), $15\fm0$ (H), and $14\fm0$ (K).  Nothing was seen in the
GLIMPSE survey of {\sl Spitzer}'s IRAC camera (Benjamin et al.~2003),
but the limits are comparable or less stringent.  We expect a surface
density of Galactic-plane X-ray sources of at least the flux of \srcx\
of about 2 deg$^{-2}$, based on the {\sl ASCA} Galactic Plane survey
(Sugizaki et al.~2001).  Within the radio error ellipse of \src, the
likelihood of finding such a source by chance is $\lapprox 10^{-4}$,
supporting its identification with \src.

The $K$-magnitude limit from 2MASS, combined with our X-ray detection,
implies an X-ray to IR flux ratio for \srcx\ of $f_x / f_K \ga
0.7$. Comparing this to the known X-ray and IR properties of stars,
galaxies and neutron stars (see, e.g., Fig.\,20 of Kaplan et
al.~2004), we find that \srcx\ has a considerably higher X-ray flux
than the bulk of the extragalactic population, and a much higher X-ray
to IR flux ratio than most stars.  On the other hand, its flux and
colors are entirely consistent with all classes of isolated neutron
star.  Deeper IR and optical observations are needed to provide better
constraints.

There is no X-ray evidence for a supernova remnant (SNR),
pulsar wind nebula or any other extended emission
anywhere within about $7'$ of \srcx.

\section{Spectral Analysis}
\label{spectrum}

We extracted the source spectrum from a circular region $15''$ in
radius centered at the X-ray source position.  Individual response
files were created for the source and for a large ($\sim3'\times 3'$)
adjacent background region.  Data from all three observations were
merged by adding spectra and combining the response files weighted by
observation times (for both source and background), and then grouped
into bins of at least 20 counts to allow the use of $\chi^2$
statistics.

\begin{figure}
\epsscale{0.80}
\plotone{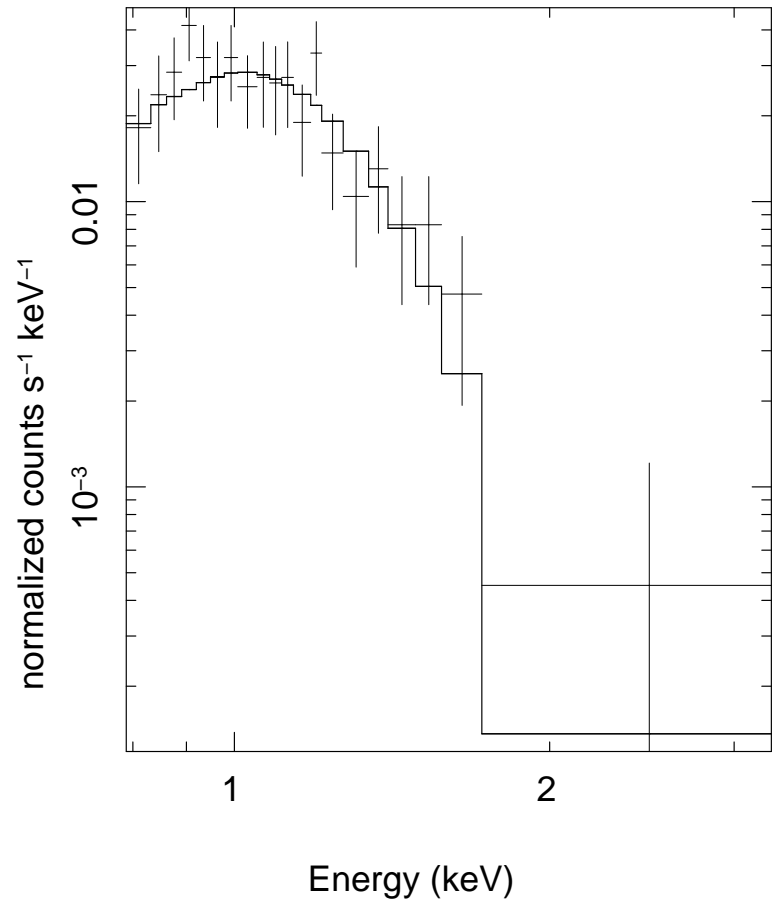}
\caption{Spectrum of \srcx, fitted with an absorbed blackbody model.
\label{fig_spectrum}}
\end{figure}

An absorbed blackbody fit (XSPEC models {\tt phabs} plus {\tt
bbodyrad}) was statistically acceptable ($\chi^2 = 10.8$ for 18
degrees of freedom); it is shown in Figure~\ref{fig_spectrum}.  An
absorbing column density $N_H = 7^{+7}_{-4} \times 10^{21}$ cm$^{-2}$
was required (all errors are 90\% confidence limits unless otherwise
specified).  The fitted temperature is $kT = 0.12 \pm 0.04$ keV and
the observed flux (0.5--8 keV) is $(1.1 \pm 0.1) \times 10^{-13}$
ergs~cm$^{-2}$ s$^{-1}$.  The absorption-corrected flux is much less
certain due to uncertainties in $N_H$: we find a flux (0.5--8 keV) of
$\sim 2 \times 10^{-12}$ ergs~cm$^{-2}$ s$^{-1}$, with uncertainty of
an order of magnitude in either direction.

A blackbody with the best-fit flux would have a radius of 10 km at a
distance of 1.8 kpc; at the DM distance of 3.6 kpc, the implied
blackbody radius is 20 km.  However, the unabsorbed flux uncertainties
are large.  Note that, given the uncertainties in DM-derived distances
(Cordes \& Lazio 2002), a distance of 1.8 kpc is still consistent with
the observed DM.  The X-ray luminosity is $3.6 \times 10^{33}(D/{3.6 \
\rm kpc})^2$ ergs~s$^{-1}$ (0.5--8 keV) (uncertain by an order of
magnitude), though of course most of the bolometric luminosity is at
lower photon energies (a spherical blackbody with $kT = 0.12$ keV and
$R = 10$ km has a total luminosity $L_{\rm bol} = 2.7 \times 10^{33}$
ergs~s$^{-1}$).  We also fit the data with a neutron-star atmosphere
model (XSPEC model {\tt nsa}; Pavlov, Shibanov, Zavlin~1991; Zavlin,
Pavlov \& Shibanov~1996), obtaining a slightly worse but acceptable
($\chi_\nu^2 = 0.66$) fit with $N_H =(0.4 - 1.3)\times 10^{22}$\,
cm$^{-2}$, $kT_{\rm eff} = 0.02 -0.2$ keV, and a flux corresponding to
roughly full-surface emission from a 10-km neutron star.  These values
are completely consistent with those of the simple blackbody fit.

An absorbed power-law fit (XSPEC models {\tt phabs} plus {\tt power})
was significantly worse ($\Delta \chi^2 = 3.9$, or worse at the 95\%
level).  In addition, the power-law fit required a photon index
$\Gamma$ of at least 9.5, far steeper than any observed magnetospheric
or other nonthermal X-ray emission from any known source.  However, a
nonthermal component at higher energies might be present in addition
to the thermal emission.  To search for such a component, we fitted
only the data above 1 keV, and obtained a best-fit power-law index of
5.5, still unreasonably steep.  There appears to be no evidence for a
nonthermal component in the X-ray spectrum of \srcx, at our current
level of sensitivity.  We note that even with the large uncertainties,
$L_x > \dot{E}$ independent of distance, making a nonthermal
explanation for most of the emission highly unlikely.

\section{Timing analysis}
\label{timing}

The time resolution of {\sl Chandra}'s CCDs in imaging mode is 3.2 s,
the read-out time.  We binned the lightcurve on a variety of time
scales to look for evidence of bursts or other time variability on
scales longer than this.

We have examined all $\sim 10^4$ of the 3.2-s CCD frames of the source
extraction region to look for any bursts resembling the radio bursts
(McLaughlin et al.~2005).  Pileup is not a concern because of the
substantially broadened off-axis PSF.  All frames contained either
zero or one events, except for 17 frames which contained two events
from the source, and one frame which contained a cluster of three
events. For the time-averaged count rate of $\sim0.05$ counts per
frame, these clusters are entirely consistent with Poisson statistics.
Given this rate, the number of events in a single frame which would
deviate from a steady flux at the 3-$\sigma$ level is 4 photons.  We
thus adopt this as the upper limit on the X-ray flux of any burst of
duration of 3.2 sec or less. Assuming a spectrum of the same shape as
that fitted for the overall source in \S\ref{spectrum} above, this
limits the observed fluence of any burst to $\la
3\times10^{-11}$~ergs~cm$^{-2}$ (0.5--8~keV), corresponding to an
absorbed flux limit of $\la 1\times10^{-8}$~ergs~cm$^{-2}$~s$^{-1}$
(0.5--8~keV) if we assume that any X-ray burst lasts for 3~ms.  At a
distance of 3.6~kpc, these limits correspond to $1\times10^{36}$~ergs
and $1\times10^{38}$~ergs~s$^{-1}$, respectively.  A burst might, of
course, have a very different spectrum from that seen in
Figure~\ref{fig_spectrum}.

We have also binned the data on longer time scales to look for any
more gradual variability. At the 3-$\sigma$ level, we find no evidence
for X-ray flux variations on any time scale ranging from 3.2~s up to
the overall time span covered of $\sim5$ days. We note however that
data were only being recorded during 7\% of this window, so within the
range of variability considered, we are not sensitive to variations on
time scales falling between $\sim 4$ hours and $\sim2$ days.

While we cannot detect pulsations at the 4.26~s period directly, we
searched for a possible alias of this period at a predicted frequency
of 0.073975824(1)\,Hz (correcting for spin-down and barycentered to
the midpoint of the three observations).  We folded the arrival times
at this frequency and found no significant power.  Including
attenuation of sensitivity at frequencies high compared to the binning
rate, our 3-$\sigma$ upper limit on the pulsed fraction $f$ is $\sim
0.5 A^{-1/2}$, where $A$ is a constant which depends on the pulse
shape (Leahy, Elsner \& Weisskopf~1983; Ransom, Eikenberry \&
Middleditch~2002).  For a narrow pulse ($A = 2$) as observed in the
radio bursts, we estimate $f \la 0.35$, while for a sinusoid ($A=0.5$)
we find $f \la 0.7$.

\section{Discussion}
\label{disc}


The spatial coincidence of \src\ and \srcx\ and the lack of any
obvious optical or infrared counterpart makes it extremely likely that
these sources are associated, and that the emitting source is a
neutron star.  We now compare the X-ray and other properties of \src\
to those seen from the other known categories of isolated neutron
star: radio pulsars, magnetars, XDINSs, and central compact objects
(CCOs).

We first note that the pulsed fraction upper limits that we obtained
in \S\ref{timing} are generally unconstraining.  Pulsed thermal
emission from a ``hot spot'' on a neutron-star surface is expected to
be broad and quasi-sinusoidal, with a low level of modulation
(Psaltis, \"Ozel \& DeDeo~2000).  Thus the pulsed fraction upper limit
of $\sim70\%$ which we have determined for a broad profile is
insufficient to have detected pulsations from almost all the sources
which we consider below.

Over much of their lives, ordinary radio pulsars emit quasi-blackbody
emission from their surfaces, the temperature dropping with time as
they cool through first neutrino, then photon emission (see Yakovlev
\& Pethick~2004 for a recent review).  A 1.4~$M_{\odot}$ neutron star
at an age of $10^4$~years is predicted by typical models to have a
surface temperature (as viewed by an observer at infinity)
$\sim100$~eV, while by $10^5$ years this has dropped to $\sim70$~eV
(Page et al.~2004; Yakovlev \& Pethick 2004).  We note that the X-ray
emission from neutron star surfaces is expected to be significantly
modified by propagation through the stellar atmosphere, which shifts
the emission into a harder component (Zavlin \& Pavlov~2002; Lloyd,
Hernquist, \& Heyl~2003).  Applied to such spectra, blackbody fits
overestimate the surface temperature by a factor of $\ga2$, and also
underestimate the radius.

While our data are insufficient for fitting to more detailed
neutron-star atmosphere models, the considerations above suggest that
the emission from \src\ is consistent with a cooling neutron star of
age\footnote{Our result that $L_x \ga \dot{E}$ for this source is not
consistent with polar-cap reheating models for the thermal X-rays seen
for much older sources (e.g. Cheng \& Zhang 1999; Harding \&
Muslimov~2001).}  $\sim10^4-10^5$~yr, at a distance $\la2$~kpc.  For
comparison, the radio pulsars B0833--45, B1706--44, J2021+3651,
J0538+2817 and B0656+14 have fitted blackbody temperatures of 130,
170, 150, 160 and 70~eV for approximate ages of 11, 17, 17, 30 and
110~kyr, respectively (Romani et al.~2005; Hessels et al.~2004; Romani
\& Ng~2003; Kramer et al.~2003; Brisken et al.~2003), again consistent
with an interpretation for \src\ as a cooling neutron star of age
$10^4-10^5$~yr.  While the temperature we find for \srcx\ is a bit
higher than that of the comparably-aged B0656+14, our errors are
large.  Better data will be required to determine whether \srcx\ is in
fact hotter than the range of predictions from standard cooling
curves.  The high magnetic-field strength might support such a
possibility (Shibanov \& Yakovlev 1996).

These crude estimates are in satisfactory agreement with the
characteristic age, $\tau_c = 117$~kyr, inferred for \src\ from its
spin-down (McLaughlin et al.~2005).  If \src\ was born spinning much
faster than its present period of 4.26~s, then for standard magnetic
dipole spin-down the characteristic age should be a good match to the
true age.  If it was born a slow rotator, then $\tau_c$ could be a
considerable overestimate (e.g., Kramer et al.~2003).

Since the surface magnetic field of \src\ is about 10 times greater
than those of the pulsars listed above, another interesting comparison
may be performed with high-$B$-field pulsars (Camilo et al.~2000;
McLaughlin et al.~2003).  The two of these sources detected in X-rays,
PSRs~J1718--3718 and J1119--6127 (Kaspi \& McLaughlin~2005; Gonzalez
et al.~2005), also show temperatures ($kT \sim 150-200$~eV) and
luminosities ($\sim10^{32}-10^{33}$~ergs~s$^{-1}$) compatible with
that of \src, although both sources are probably much younger (35~kyr
and 1.7~kyr, respectively), and have $L_x < \dot{E}$.


The high inferred surface magnetic field strength, long spin period,
and lack of persistent radio emission also suggest a comparison of
\src\ with the population of magnetars. These objects (which include
both the anomalous X-ray pulsars and soft gamma repeaters) are
characterized by quiescent, bursting, and flaring X-ray emission all
powered in different ways by ultra-strong magnetic fields (Woods \&
Thompson 2006, and references therein).  However, magnetars are
typically hotter ($kT \sim 0.3-0.6$ keV), show a nonthermal spectral
component with $\Gamma \sim 2 - 4$, and are brighter by one to three
orders of magnitude ($L_X \sim 10^{34} - 10^{36}$ ergs s$^{-1}$; Woods
\& Thompson~2006). Moreover, the characteristics of the radio bursts
seen from \src\ and from the other RRATs are completely different in
their energetics and recurrence times from the much rarer X-ray bursts
seen from the magnetars.  The soft X-ray spectrum of \src\ does have a
comparable temperature to the quiescent state of the transient
magnetar XTE~J1810--197 ($kT \sim 0.15-0.18$ keV; Ibrahim et al.~2004;
Gotthelf et al.~2004).  However, the X-ray luminosity of the latter is
$\sim10$ times larger. XTE~J1810--197 also has a possible transient
radio counterpart (Halpern et al.~2005), though with quite different
properties from the RRATs.  Further observations are required for a
detailed comparison with \src.


The lack of persistent radio emission and the long spin period of
\src\ also raise the possibility of a link with the XDINSs
(Haberl~2004).  The XDINSs are slightly cooler ($kT \sim
0.04-0.1$\,keV) and less luminous ($L_X \sim 10^{31}-10^{32}$\,ergs
s$^{-1}$) than \src.  However, the measured period derivatives of two
XDINSs (RBS~1223 and RX~J0720.4--3125; Kaplan \& van~Kerkwijk~2005a,
2005b), and the detection of possible proton cyclotron lines in their
spectra (van~Kerkwijk~2004), imply magnetic field strengths similar to
those of \src. No radio emission of any kind has been reported from
XDINSs; recent observations (Bradley 2006) show no RRAT-like radio
bursts toward RX~J0720.4--3125 (or toward magnetars).

\src\ does not appear to have an associated SNR.  It has no apparent
connection to the nearby SNR G15.9+0.2 (see below), and shows no
extended radio or X-ray emission around its position.  Nevertheless,
for completeness we compare its X-ray emission to that seen from the
so-called CCOs, a small and heterogeneous sample of X-ray point
sources seen in young SNRs, corresponding to a population of young
neutron stars whose connection to pulsars, magnetars and XDINSs is as
yet unclear (see Pavlov, Sanwal \& Teter~2004 for a review).  The CCOs
have spectra which can generally be fitted by a blackbody plus a
power-law tail, the former component having a temperature $kT \approx
0.3-0.5$~keV with a typical bolometric luminosity $L_{\rm bol} \ga
10^{33}$~ergs~s$^{-1}$.  Periodicities have been reported for four of
the CCOs, but with periods much faster (100--400~ms) or much slower
($\sim6$~hr) than seen for \src.  The properties of \src\ thus seem to
have little in common with those seen from the CCOs, although \src\ is
likely much older than these objects and the evolutionary paths
followed by the CCOs are unclear in any case.

The shell SNR G15.9+0.2 has a much higher absorbing column density
($\sim 4 \times 10^{22}$ cm$^{-2}$) than \srcx, suggesting that it is
at considerably greater distance.  In any case, G15.9+0.2 appears to
be younger than 1000 yr (Reynolds et al., in preparation), so
disregarding the much older spin-down age of \src\ and hypothesizing
an association, the $10\farcm3$ offset between \src\ and the center of
G15.9+0.2 requires a tranverse velocity for the former of
$>10\,000$~km~s$^{-1}$ at a distance of 3.6~kpc.  This makes any
physical association of G15.9+0.2 with \src\ highly improbable.

\section{Conclusions}

We have discovered the first X-ray counterpart to a Rotating RAdio
Transient source, \srcx.  The X-ray source is well described by a
thermal spectrum consistent with emission from a cooling neutron star
of age $10^4-10^5$~years, broadly consistent with the characteristic
age of \src.  The X-ray properties also suggest possible connections
to the population of X-ray dim isolated neutron stars and to the
transient magnetar XTE J1810--197. A search for an X-ray modulation at
the aliased radio pulse frequency was unsuccessful. No variations were
seen in the X-ray flux on longer time scales either, and no optical
and infrared counterparts to the source have been found.  Deeper X-ray
observations are required to search for pulsations, bursts, and in
general to clarify the nature of the RRATs.

\acknowledgments

This work was supported by NASA through Chandra General Observer
Program grant GO5-6051A. B.M.G.~acknowledges the support of NASA
through LTSA grant NAG5-13023, and of an Alfred P.\ Sloan Research
Fellowship. A.P.~and M.B.~received support from the Italian Ministry
of University and Research under the national program PRIN 2003.  We
thank the astronomers involved in maintaining the VizieR browser and
the 2MASS catalogue.

\newpage





\end{document}